\documentclass[10pt,aps,prd,floatfix,notitlepage,showpacs,nofootinbib,superscriptaddress,showkeys]{revtex4-1}
\usepackage{amssymb,amsmath}
\usepackage[colorlinks=magenta, colorlinks, linkcolor=blue, citecolor=magenta]{hyperref}
\usepackage{graphicx}

\begin{document}
	
	
	\title{Gravitational energy-momentum pseudo-tensor in  Palatini and metric  $f(R)$ gravity}
	
	
	\author{Habib Abedi}
	\email{h.abedi@ut.ac.ir}
	\affiliation{Department of Physics, University of Tehran, North Kargar Avenue, 14399-55961 Tehran, Iran}
	\author{Salvatore Capozziello}
	\email{capozziello@unina.it}
	\affiliation{Dipartimento di Fisica "E. Pancini", Universit\`a di Napoli {}``Federico II'', Compl. Univ. di
		   Monte S. Angelo, Edificio G, Via Cinthia, I-80126, Napoli, Italy, }
		  \affiliation{Istituto Nazionale di Fisica Nucleare, Sezione  di Napoli, Compl. Univ. di
		   Monte S. Angelo, Edificio G, Via Cinthia, I-80126,  Napoli, Italy,}
		   \affiliation{Scuola Superiore Meridionale, Largo S. Marcellino 10, I-80138, Napoli, Italy,}
 \affiliation{Department of Mathematics, Faculty of Civil Engineering,
VSB-Technical University of Ostrava, Ludvika Podeste 1875/17, 
708 00 Ostrava-Poruba, Czech Republic.}
 
        \author{Maurizio Capriolo}
	\email{mcapriolo@unisa.it}
	 \affiliation{Istituto Nazionale di Fisica Nucleare, Sezione  di Napoli, Compl. Univ. di
		   Monte S. Angelo, Edificio G, Via Cinthia, I-80126,  Napoli, Italy,}		   
		   \affiliation{Dipartimento di Matematica,  Universit\`a di Salerno, via Giovanni Paolo II, 132, Fisciano, SA I-84084, Italy}
	\author{Amir M. Abbassi}
	\email{amabasi@khayam.ut.ac.ir}
	\affiliation{Department of Physics, University of Tehran, North Kargar Avenue, 14399-55961 Tehran, Iran}
	\begin{abstract}
		We derive the gravitational energy-momentum pseudo-tensor $\tau^\mu_{\phantom{\mu}\nu}$ in both Palatini and metric  approaches to  $f(R)$ gravity. We then obtain the related cosmological gravitational energy density. Considering a flat Friedmann-Lema\^itre-Robertson-Walker  spacetime, the energy density complex of matter and gravitation vanishes in  the metric approach, but results   non-vanishing  in the Palatini formalism. This feature could be relevant in order to physically discriminate between the two approaches.
			\end{abstract}
	\date{\today}
	\maketitle
	\section{Introduction}
	The gravitational energy and its localization in curved spacetime  is one of the oldest problems in General Relativity (GR). The issue is related to the fact that  the matter energy-momentum tensor is not locally conserved  in curved spacetime, i.e. $T_{\mu\phantom{\nu},\nu}^{\nu}\neq 0$. Einstein was the first who introduced locally conserved energy-momentum pseudo-tensor including both the gravitational and matter energy-momentum contributions~\cite{Einstein}. Several attempts have been made to introduce prescriptions for the energy-momentum complex showing advantages and disadvantages (see~\cite{Hasten} and references therein for a comprehensive review).  However,   a generally accepted definition for localized gravitational energy does not exist.  
For example, in cosmology, it has been shown that the energy complex leads to  vanishing values for open and closed  Friedmann-Lema\^itre-Robertson-Walker (FLRW) spacetimes ~\cite{Rosen:1994vj, Vargas:2003ak, Johri:1995gh, Faraoni:2002vw, Abedi:2018eyz} and, in general,  the problem of a correct  definition of energy complex could be strictly related to the dark side issue. In this perspective, also modified gravity theories contribute to the debate in the sense that dynamical  effects of  dark  components could be addressed by geometrical extensions or modifications of GR without invoking new particle counterparts, beyond the Standard Model, up to now not detected at fundamental level. 
	
A number of different alternative theories of gravity have been proposed. One of the most popular  is $f(R)$ gravity  where the  Hilbert-Einstein action, linear in the Ricci curvature scalar $R$, is substituted by a general function $f(R)$ \cite{Cap1, Rept,Odi1,Odi2}. 	This straightforward extension of GR led also to a fundamental discussion on the effective degrees of freedom and physical variables of gravitational field. 
In GR, the underlying assumption of the geometric structures on spacetime is that the affine connection is the Levi-Civita connection of the metric. However, a purely metric formulation is not the only way to construct a theory of gravity. The Palatini approach, where the metric and the affine connection are regarded as independent variables (metric-affine formalism),  yields interesting results questioning the central role of metric $g_{\mu\nu}$ to describe the gravitational field.  The intriguing issue is that this approach results, in the  spacetime structure and  field equations, exactly as GR if formulated  for the Einstein-Hilbert action. However,  for  more general Lagrangian densities, like higher-order or scalar tensor theories, the purely metric or  Palatini formulations result very different.   In particular, the  Palatini $f({\cal R})$ gravity leads to different field equations from  metric  $f(R)$ gravity\footnote{We will indicate with ${\cal R}$ the curvature scalar in metric-affine formalism and with $R$ the same quantity in metric formalism. This notation is adopted also for the other curvature invariants.}~\cite{Olmo:2011uz,Mauro}.  

In the present work, we will consider the gravitational energy density complex  in $f(R)$ gravity formulated in Palatini and metric approaches. In particular, we will focus on the cosmological aspects.  
	
This discussion enters into  the problem of energy-momentum localization recently considered  in some modified theories of gravity like $f(R)$ itself,  ${f(R,\Box R,\dots, \Box^{k} R)}$ ~\cite{Multamaki:2007wb, Capozziello:2018qcp}, teleparallel gravity and its extended version $f(T)$~\cite{Capozziello:2017xla, Abedi:2015cya}.	

	The paper is organized as follows. In Sec.~\ref{paltinisection}, we derive the field equations and the gravitational energy-momentum pseudo-tensor in Palatini $f({\cal R})$ gravity. Considering  a flat FLRW spacetime, we explicitly calculate the energy density for cosmological power law expansions. Sec.~\ref{metricsection} is devoted to the gravitation pseudo-tensor in metric formalism of $f(R)$. Also in this case,   cosmological  power law  solutions are considered.   Sec.\ref{conclusions} is devoted to conclusions.
	
	\section{The Palatini formulation of $f(R)$ gravity}
	\label{paltinisection}
	The Palatini action of $f({\cal R})$ is given by
	\begin{align} \label{actionf}
		{\cal S} = \frac{1}{2 \kappa^2} \int {\rm d}^4x \,\sqrt{-g}\, f({\cal R}),
	\end{align}
	with $\kappa^2=8\pi G /c^4$, $g$ the determinant of metric tensor $g_{\mu\nu}$ and the Riemann and Ricci tensors defined as 
	\begin{align}
		{\cal R}_{\mu\nu}(\Gamma)=& \partial_\alpha\Gamma^\alpha_{\mu\nu}-\partial_\nu\Gamma^\alpha_{\mu\alpha}+\Gamma^\alpha_{\mu\nu}\,\Gamma^\sigma_{\alpha\sigma} -\Gamma^\alpha_{\nu\lambda} \, \Gamma^\lambda_{\mu\alpha},
		\\
		{\cal R}(g,\Gamma) =& {\cal R}_{\mu\nu}(\Gamma)\, g^{\mu \nu}.
	\end{align} 
	We use the metric $g_{\mu\nu}$ and the connection $\Gamma^\alpha_{\mu\nu}$ as independent variables which means that we  do not assume any relation between the metric and the connection.
	The variation of the metric $g^{\mu\nu}$, the connection $\Gamma^\alpha_{\mu\nu}$, and coordinate~$x^\mu$ for a general infinitesimal transformation yields
	\begin{align}
		x^{\prime \mu}  =& x^\mu + \delta x^\mu,
		\\
		g^{\prime \mu\nu}(x^\prime) =& g^{\mu\nu}(x) + \tilde{\delta}g^{\mu\nu} ,&
		g^{\prime \mu\nu}(x) =& g^{\mu\nu}(x) + \delta g^{\mu\nu},
		\\
		\Gamma^{\prime \alpha}_{\mu\nu}(x^\prime) =& \Gamma^\alpha_{\mu\nu}(x) + \tilde{\delta} \Gamma^\alpha_{\mu\nu},&
		\Gamma^{\prime\alpha}_{\mu\nu}(x) =& \Gamma^\alpha_{\mu\nu}(x) + \delta \Gamma^\alpha_{\mu\nu},
	\end{align}
	where $\tilde{\delta}$ is the local variation and $\delta$ is the variation for fixed coordinate $x$.
	Variation of the action with respect to the metric $g^{\mu\nu}$,  the connection $\Gamma^{\alpha}_{\beta,\gamma}$ , and the coordinates $x^\mu$ results in
	\begin{align}
		\tilde{\delta} {\cal S} = \frac{1}{2\kappa^2} \int {\rm d}^4x \Bigg\{ \sqrt{-g}\left[\left(f_{\cal R} {\cal R}_{\mu\nu}-\frac{1}{2} g_{\mu\nu} f\right) \, \delta g^{\mu\nu} + f_{\cal R} g^{\mu\nu} \, \delta {\cal R}_{\mu\nu}\right] +\partial_\mu \left(\sqrt{-g} f \, \delta x^\mu\right) \Bigg\},
	\end{align}
	where $f_{\cal R}:={\rm d} f({\cal R})/{\rm d}{\cal R}$. This can be simplified by the Palatini identity
	\begin{align}
		\delta {\cal R}_{\mu\nu} = \nabla_\alpha\left(\delta \Gamma^\alpha_{\mu\nu}\right)-\nabla_\nu\left(\delta\Gamma^\alpha_{\alpha\mu}\right).
	\end{align}
	Then, we get
	\begin{align} \label{action1}
		\tilde{\delta} {\cal S} = \frac{1}{2\kappa^2 } \int {\rm d}^4x \, \Bigg\{ & \sqrt{-g}\left(f_{\cal R} {\cal R}_{\mu\nu}-\frac{1}{2} g_{\mu\nu} f\right) \, \delta g^{\mu\nu} 
		+\delta \Gamma^\lambda_{\phantom{\lambda}\nu\mu}\left[-\nabla_\lambda\left(\sqrt{-g}g^{\nu\mu}f_{\cal R}\right)+\nabla_\alpha\left(\sqrt{-g}g^{\mu\alpha} \delta^\nu_\lambda f_{\cal R}\right)\right]
		\nonumber \\&
		+
		\partial_\lambda\left[\sqrt{-g}f_{\cal R} \left(g^{\mu\nu} \delta^\lambda_\alpha-g^{\mu\lambda} \delta^{\nu}_\alpha \right)  \delta \Gamma^\alpha_{\phantom{\alpha}\mu\nu} +\sqrt{-g} f \, \delta x^\lambda \right]
		\Bigg\}.
	\end{align}
	We consider the action  stationary at a fixed point $x$ in a given domain of the action~\eqref{actionf}. This means that the variation of  metric and its first derivatives vanish at the boundary.  The field equations for the metric and the connection in vacuum  are
	\begin{align}\label{FieldEq1}
		f_{\cal R} {\cal R}_{(\mu\nu)}-\frac{1}{2} g_{\mu\nu} f =& 0,
		\\ \label{FieldEq2}
		\nabla_\lambda\left(\sqrt{-g}g^{\nu\mu}f_{\cal R}\right) =& 0,
	\end{align}
	where ${\cal R}_{(\mu\nu)}$ is the symmetric part of the Ricci tensor because $\cal R_{\mu\nu}$ loses its symmetry using an arbitrary non-metric compatible connection that is
	\begin{equation}
	\cal R_{\mu\nu}=\cal R_{\nu\mu}+\cal R^{\lambda}_{\phantom{\lambda}\lambda\mu\nu}.
	\end{equation}
	Given that the Riemann tensor $\cal R^{\sigma}_{\phantom{\sigma}\lambda\mu\nu}$ is no longer antisymmetric on its first two indices,  then the term $\cal R^{\lambda}_{\phantom{\lambda}\lambda\mu\nu}$ is not equal to zero.
	For a generic infinitesimal transformation, we have
	\begin{align}
		x^{\prime \mu} =& \, x^\mu + \xi^\mu,
		\\
		g^{\prime \mu\nu}(x^\lambda) \simeq& \,  g^{\mu\nu}(x^\lambda) - \xi^\lambda \frac{\partial g^{\mu\nu}}{\partial x^\lambda} + g^{\mu\alpha} \frac{\partial \xi^\nu}{\partial x^\alpha} + g^{\nu\alpha} \frac{\partial \xi^\mu}{\partial x^\alpha},
		\\
		\Gamma^{\prime \alpha}_{\phantom{\alpha}\mu\nu} (x^\lambda) \simeq& \, \Gamma^{\alpha}_{\phantom{\alpha}\mu\nu} (x^\lambda) - \xi^\lambda \frac{\partial \Gamma^{\alpha}_{\phantom{\alpha}\mu\nu}}{\partial x^\lambda} + \Gamma^{\rho}_{\phantom{\rho}\mu\nu} \frac{\partial \xi^\alpha}{\partial x^\rho} -\Gamma^{\alpha}_{\phantom{\alpha}\sigma\nu} \frac{\partial \xi^\sigma}{\partial x^\mu} -\Gamma^{\alpha}_{\phantom{\alpha}\mu\sigma} \frac{\partial \xi^\sigma}{\partial x^\nu} - \frac{\partial^2 \xi^\alpha}{\partial x^\mu \, \partial x^\nu},
	\end{align}
	where we have neglected terms of higher order in $\xi^\mu$ in the series expansion.
	For an infinitesimal transformation like a rigid translation with $\partial_\mu \xi^\nu =0$, we get
	\begin{align}
		g^{\prime \mu\nu}(x^\lambda) \simeq \, &   g^{\mu\nu}(x^\lambda) - \xi^\lambda \frac{\partial g^{\mu\nu}}{\partial x^\lambda},
		\\
		\Gamma^{\prime \alpha}_{\phantom{\alpha}\mu\nu} (x^\lambda) \simeq \, &  \Gamma^{\alpha}_{\phantom{\alpha}\mu\nu} (x^\lambda) - \xi^\lambda \frac{\partial \Gamma^{\alpha}_{\phantom{\alpha}\mu\nu}}{\partial x^\lambda} .
	\end{align}
	Then, the action~\eqref{action1} becomes
	\begin{align}
		\tilde{\delta} {\cal S}_{\rm g} =& \frac{1}{2\kappa^2} \int {\rm d}^4x \Bigg\{ - \sqrt{-g}\left(f_{\cal R} {\cal R}_{\mu\nu}-\frac{1}{2} g_{\mu\nu} f\right) \, \xi^\lambda \, g^{\mu\nu}_{\phantom{\mu\nu},\lambda}
		-\xi^\lambda \, \Gamma^\beta_{\phantom{\beta}\nu\mu,\lambda}\left[-\nabla_\beta\left(\sqrt{-g}g^{\nu\mu}f_{\cal R}\right)+\nabla_\alpha\left(\sqrt{-g}g^{\mu\alpha} \delta^\nu_\beta f_{\cal R}\right)\right]
		\nonumber \\&
		+
		\partial_\lambda\left[-\sqrt{-g}f_{\cal R} \left(g^{\mu\nu} \delta^\lambda_\alpha-g^{\mu\lambda} \delta^\nu_\alpha \right) \, \xi^\beta \, \Gamma^\alpha_{\phantom{\alpha}\mu\nu,\beta} +\sqrt{-g} f \, \xi^\lambda \right]
		\Bigg\}.
	\end{align}
	Considering the local variation of the action, we have
	\begin{align}
		\partial_\lambda\left(\sqrt{-g} \, \tau^\lambda_{\phantom{\lambda}\beta} \right) = 0,
	\end{align}
	where the gravitational energy-momentum  pseudo-tensor $\tau^\lambda_{\phantom{\lambda}\beta}$ of $f({\cal R})$ gravity  is 
	\begin{align}
		\tau^\lambda_{\phantom{\lambda}\beta}=\frac{1}{2\kappa^{2}} \left[f\left({\cal R}\right) \, \delta^\lambda_\beta -f_{\cal R}\left({\cal R}\right)\, \left(g^{\mu\nu} \delta^\lambda_\alpha-g^{\mu\lambda} \delta^\nu_\alpha \right)  \, \Gamma^\alpha_{\phantom{\alpha}\mu\nu,\beta} \right].
	\end{align}
	Let us consider now the action containing the matter part, that is 
	\begin{align}
		{\cal S}_{\rm m} = \int {\rm d}^4x \, \sqrt{-g}\,{\cal L}_{\rm m}.
	\end{align}
	In the general case, the matter Lagrangian ${\cal L}_{\rm m}$ depends on the connection that, for example, comes naturally in presence of fermion fields.
	Here, we  consider only  material Lagrangian which does not depend on the affine connection $\Gamma$. Then the matter energy-momentum tensor is defined by
	\begin{align} \label{mEM}
		T_{\mu\nu} =& -\frac{2}{\sqrt{-g}}\frac{\delta\left(\sqrt{-g}{\cal L}_{\rm m}\right)}{\delta g^{\mu\nu}}.
	\end{align}
	Field equations for  metric and  connection, i.e. Eqs.~\eqref{FieldEq1} and~\eqref{FieldEq2}, in presence of  matter  become
	\begin{align}\label{Einstein and Palatini Eqs.}
		f_{\cal R} {\cal R}_{(\mu\nu)}-\frac{1}{2} g_{\mu\nu} f =& \kappa^2 T_{\mu\nu},
		\\
		\nabla_\lambda\left(\sqrt{-g}g^{\nu\mu}f_{\cal R}\right) =& 0 \label{bimetric}.
	\end{align}
	The connection, in general,  is  not metric compatible, i.e.  $\nabla_{\lambda}g_{\mu\nu}\neq 0$.
	In this formalism, one can define a new metric conformally related to the metric $g_{\mu\nu}$,
	\begin{align}
		h_{\mu\nu}:=f_{\cal R}g_{\mu\nu}.
	\end{align}
	Then Eq.~\eqref{bimetric}  reduces to 
	\begin{align}
		\nabla_\lambda \left(\sqrt{h}h^{\mu\nu}\right)=0.
	\end{align}
	Then  connection $\Gamma^\alpha_{\phantom{\alpha}\mu\nu} $ results the Christoffel symbol constructed by  $h_{\mu\nu}$, i.e. \begin{align}\label{PalatiniConnection}
		\Gamma^\alpha_{\phantom{\alpha}\mu\nu}=
		\frac{1}{2\, f_{\cal R}\left({\cal R}\right)}g^{\alpha \beta} \left[\partial_\mu\left( f_{\cal R}\left({\cal R}\right) g_{\nu\beta}\right)+\partial_\nu\left( f_{\cal R}\left({\cal R}\right) g_{\mu\beta}\right)-\partial_\beta\left( f_{\cal R}\left({\cal R}\right) g_{\mu\nu}\right)\right]\ .
	\end{align}
	The connection $\Gamma^\alpha_{\phantom{\alpha}\mu\nu} $ is related to the Levi-Civita connection $\stackrel{\circ}{\Gamma}{}^\alpha_{\phantom{\alpha}\mu\nu} $ by the expression
	\begin{equation}
	\Gamma^\alpha_{\phantom{\alpha}\mu\nu}=\,\stackrel{\circ}{\Gamma^{\alpha}}_{\mu\nu}+\delta^{\alpha}_{\mu}A_{\nu}+\delta^{\alpha}_{\nu}A_{\mu}-g_{\mu\nu}A^{\alpha}\ ,
	\end{equation}
	where the four-vector $A_{\mu}$ is defined as
	\begin{equation}
	A_{\mu}:=\frac{1}{2f_{\cal R}}\nabla_{\mu}f_{\cal R}\ .
	\end{equation}
	For $f({\cal R})={\cal R}$, we recover  the Christoffel symbols constructed by the metric $g_{\mu\nu}$, that is 
	\begin{align}
	\Gamma^\alpha_{\phantom{\alpha}\mu\nu} = \frac{1}{2}g^{\alpha\beta}\left( g_{\beta\mu,\nu}+g_{\beta\nu,\mu}-g_{\mu\nu,\beta}\right)\,,
	\end{align}
	this means that  the standard Levi-Civita connection is recovered and then no difference results in metric and Palatini formalism for GR. 
	
	The Ricci tensor ${\cal R}_{\mu\nu}$ can be written as follows
	\begin{align} \label{Rconf}
		{\cal R}_{\mu\nu} = & R_{\mu\nu} +\frac{3}{2} \frac{1}{\left(f_{\cal R}({\cal R})\right)^2} \left(\stackrel{\circ}{\nabla}_\mu f_{\cal R}({\cal R}) \right) \, \left(\stackrel{\circ}{\nabla}_\mu f_{\cal R}({\cal R})\right) - \frac{1}{ f_{\cal R}({\cal R})} \left( \stackrel{\circ}{\nabla}_\mu \stackrel{\circ}{\nabla}_\nu-\frac{1}{2}g_{\mu\nu} \stackrel{\circ}{\square}\right) f_{\cal R}({\cal R}),
	\end{align}
	where $\stackrel{\circ}{\square} := \stackrel{\circ}{\nabla}{}^\mu \stackrel{\circ}{\nabla}_\mu$ and $\stackrel{\circ}{\nabla}$ denotes the covariant derivative associated to the Levi-Civita connection. Contracting Eq.~\eqref{Rconf} with $g^{\mu\nu}$ gives
	\begin{align}
		{\cal R} = R + \frac{3}{2\left(f_{\cal R}({\cal R})\right)^2} \left(\stackrel{\circ}{\nabla}_\mu f_{\cal R}({\cal R}) \right)\; \left(\stackrel{\circ}{\nabla}{ }^\mu f_{\cal R}({\cal R})\right) + \frac{3}{ f_{\cal R}({\cal R})} \stackrel{\circ}{\square}   f_{\cal R}({\cal R}).
	\end{align}
	Using the connection $\Gamma^\alpha_{\phantom{\alpha}\mu\nu}$, the symmetry of Ricci tensor is restored because the relation
	\begin{equation}
	\Gamma_{\lambda}=\frac{\partial_{\lambda}{\left(f_{\cal R}^{2}\sqrt{-g}\right)}}{f_{\cal R}^{2}\sqrt{-g}},
	\end{equation}
	gives
	\begin{equation}
	{\cal R}_{[\mu\nu]}=\partial_{[\mu}\Gamma_{\nu]}=0.
	\end{equation}
	On the other hand, the connection is not metric compatible with respect to $g_{\mu\nu}$ being
	\begin{equation}
	\nabla_{\lambda}g_{\mu\nu}=-\frac{g_{\mu\nu}}{f_{\cal R}}\nabla_{\lambda}f_{\cal R}.
	\end{equation}
	However,   the following property for the commutator 
	\begin{equation}
	\left[\nabla_{\rho},\nabla_{\lambda}\right]g_{\mu\nu}=0\,,
	\end{equation}
	holds.
	Thus, by the definition of  Riemann tensor for an arbitrary tensor $J_{\mu\nu}$, we have
	\begin{equation}
	\left[\nabla_{\rho},\nabla_{\lambda}\right]J_{\mu\nu}=-{\cal R}^{\alpha}_{\phantom{\alpha}\mu\rho\lambda}J_{\alpha\nu}-{\cal R}^{\alpha}_{\phantom{\alpha}\nu\rho\lambda}J_{\mu\alpha}.
	\end{equation}
	We obtain the antisymmetry on the first two indices of Riemann tensor,  namely
	\begin{equation}
	\cal R_{\mu\nu\lambda\rho}=-\cal R_{\nu\mu\lambda\rho}.
	\end{equation}
	It is then possible to verify that the contracted Bianchi identity holds, that is 
	\begin{equation}
	\nabla_{\mu}\left({\cal R}^{\mu\nu}-\frac{1}{2}g^{\mu\nu}{\cal R}\right)=0.
	\end{equation}
	By the Palatini connection Eq.\eqref{PalatiniConnection} and symmetry of energy-momentum tensor $T_{\mu\nu}$, having defined the new metric $h_{\mu\nu}$ and taking into account that
	\begin{align}
		\Gamma_{\lambda}=\frac{\partial_{\lambda}{\sqrt{-h}}}{\sqrt{-h}}
	\end{align}
	and 
	\begin{align}
		\Gamma_{\mu\nu\lambda}+\Gamma_{\nu\mu\lambda}=\frac{1}{f_{{\cal R}}}\partial_{\lambda}h_{\mu\nu},
	\end{align}
	we obtain
	\begin{align}\label{50}
		\sqrt{-h}\nabla_{\sigma}T^{\sigma}_{\phantom{\sigma}\nu}=\partial_{\sigma}\left(\sqrt{-h}T^{\sigma}_{\phantom{\sigma}\nu}\right)-\frac{1}{2f_{\cal R}}T^{\lambda\rho}\partial_{\nu}h_{\lambda\rho}\sqrt{-h}.
	\end{align}
	In presence of matter, we have, from Eqs.~\eqref{Einstein and Palatini Eqs.},
	\begin{align}\label{55}
		0=\frac{\sqrt{-h}}{2f_{\cal R}^{2}}T^{\mu\nu}g_{\mu\nu,\beta}\xi^{\beta}+\partial_{\lambda}\left\{\sqrt{-g}\frac{1}{2\kappa^{2}}\left[
		f\left({\cal R}\right) \, \delta^\lambda_\beta -f_{\cal R}\left(g^{\mu\nu} \delta^\lambda_\alpha-g^{\mu\lambda} \delta^\nu_\alpha \right)  \, \Gamma^\alpha_{\phantom{\alpha}\mu\nu,\beta} \right]\xi^{\beta}\right\}.
	\end{align}
	From Eq.~\eqref{50}, after some algebraic manipulations, we get 
	\begin{align}\label{56}
		\partial_{\sigma}\left[\sqrt{-g}\left(T^{\sigma}_{\phantom{\sigma}\beta}+t^{\sigma}_{\phantom{\sigma}\beta}\right)\right]=\frac{\sqrt{-h}}{f_{\cal R}^2} \, \nabla_{\lambda}T^{\lambda}_{\phantom{\lambda}\beta}+\frac{2\sqrt{-h}}{f_{\cal R}^{3}}T^{\lambda}_{\phantom{\lambda}\beta} \, \nabla_{\lambda}f_{\cal R}-\frac{\sqrt{-h}}{2f_{\cal R}^{3}}T \, \nabla_{\beta}f_{\cal R}.
	\end{align}
	Considering the contracted Bianchi identity and the field equations, the following relations
	\begin{equation}
	\left[\nabla_{\mu},\nabla_{\nu}\right]\nabla^{\mu}f_{\cal R}={\cal R}^{\alpha}_{\phantom{\alpha}\nu}\nabla_{\alpha}f_{\cal R},
	\end{equation}
	and
	\begin{equation}\label{60}
	\kappa^{2}\nabla_{\mu}T^{\mu}_{\phantom{\mu}\nu}={\cal R}^{\alpha}_{\phantom{\alpha}\nu}\nabla_{\alpha}f_{\cal R}=\left[\nabla_{\mu},\nabla_{\nu}\right]\nabla^{\mu}f_{\cal R},
	\end{equation}
	hold.
	The trace  of Eqs.~\eqref{Einstein and Palatini Eqs.} gives the so called   {\it structural equation} of space-time \cite{Allemandi}, that is 
	\begin{equation}\label{65}
	T=\frac{1}{\kappa^{2}}\left[f_{\cal R}{\cal R}-2f\left({\cal R}\right)\right],
	\end{equation}
	where $T=T_{\mu\nu}g^{\mu\nu}$.  For a given $f({\cal R})$,  we can, in principle, solve this equation and get a relation ${\cal R}={\cal R}(T)$. Thanks  to Eq.~\eqref{65}, considering $T=0$, the theory reduces to GR with a cosmological constant.

	Inserting  Eqs.~\eqref{60} and~\eqref{65} into Eq.~\eqref{56}, we obtain 
	\begin{equation} \label{47}
	\partial_{\sigma}\left[\sqrt{-g}\left(T^{\sigma}_{\phantom{\sigma}\beta}+\tau^{\sigma}_{\phantom{\sigma}\beta}\right)\right]=-\frac{\sqrt{-g}}{\kappa^{2}}G^{\lambda}_{\phantom{\lambda}\beta}\nabla_{\lambda}f_{\cal R},
	\end{equation}
	where $G^{\lambda}_{\phantom{\lambda}\beta}$ is the Einstein tensor.
	After some algebraic manipulations, we obtain the following expression 
	\begin{equation}\label{61}
	G^{\lambda}_{\phantom{\lambda}\beta}\nabla_{\lambda}f_{\cal R}=-\kappa^{2}\stackrel{\circ}{\nabla}_{\mu}T^{\mu}_{\phantom{\mu}\beta}.
	\end{equation}
	The right hand side of Eq.~\eqref{61} vanishes~\cite{Dick:1992jn, Barraco:1998eq, Koivisto:2005yk}, and then, according to Eqs.~\eqref{47} and~\eqref{61}, the energy-momentum  complex for Palatini $f({\cal R})$ gravity  is conserved, namely
	\begin{equation}
	\partial_{\sigma}\left[\sqrt{-g}\left(T^{\sigma}_{\phantom{\sigma}\beta}+\tau^{\sigma}_{\phantom{\sigma}\beta}\right)\right]=0.
	\end{equation}

As a concrete example of the above considerations,  let us take into account a  flat FLRW spacetime whose  line element is 
	\begin{align}
		{\rm d}s^2= -{\rm d}t^2+ a^2(t)\, \left({\rm d}x^2+{\rm d}y^2+{\rm d}z^2\right),
	\end{align}
	where $ a(t) $ is the scale factor of the universe and $t$  the cosmic time.  From the relation~\eqref{Rconf} and the field equations \eqref{Einstein and Palatini Eqs.}, we get
	\begin{align}
		2\kappa^2 T^0_{\phantom{0}0} = & -f +6f_{\cal R} \left(\dot{H}+H^2\right) +\ddot{f}_{\cal R} -3\frac{\dot{f}_{\cal R}^2}{f_{\cal R}} -3H\dot{f}_{\cal R},
	\end{align}
	where $H=\dot{a}/a$ is the Hubble parameter and dots denote the derivatives with respect to the cosmic time $t$. The $\tau^0_{\phantom{0}0}$ is
	\begin{align}
		2\kappa^2 \tau^0_{\phantom{0}0} = & f -6f_{\cal R} \left(\dot{H}+H^2\right) -3\ddot{f}_{\cal R} +3\frac{\dot{f}_{\cal R}^2}{f_{\cal R}} -3H\dot{f}_{\cal R}.
	\end{align}
	Then,
	\begin{align}
		\kappa^2(\tau^0_{\phantom{0}0}+T^0_{\phantom{0}0}) = -\ddot{f}_{\cal R} -3H \dot{f}_{\cal R}.
	\end{align} 
	In the case of GR, i.e. $f({\cal R})={\cal R}$, we get 
	\begin{align}
	\left(t^0_{\phantom{0}0}+T^0_{\phantom{0}0}\right)\Big|_{\rm GR}=0\,.
	\end{align}
	Let us now assume  that matter is described by  perfect fluids  including radiation and non-relativistic dust with the following energy-momentum tensor for each components, respectively,
	\begin{align}
		\left(T^\mu_{\phantom{\mu}\nu} \right)_{\rm r} =& {\rm diag}\left(-\rho_{\rm r},p_{\rm r},p_{\rm r},p_{\rm r} \right) & \mbox{ with equation of state}& &p_{\rm r}=&\frac{1}{3} \rho_{\rm r},
		\\
		\left(T^\mu_{\phantom{\mu}\nu} \right)_{\rm m} =& {\rm diag}\left(-\rho_{\rm m},p_{\rm m},p_{\rm m},p_{\rm m} \right), & \mbox{ with equation of state }& &p_{\rm m}=&0,
	\end{align}
	where $\rho_{\rm i}$ and $p_{\rm i}$ are the energy density and pressure of each fluid component.
	From the conservation of energy-momentum tensor, we get, respectively, 
	\begin{align} \label{c1}
		\dot{\rho}_{\rm r}+4 H \rho_{\rm r}=&0,
		\\ \label{c2}
		\dot{\rho}_{\rm m}+3 H \rho_{\rm m}=&0.
	\end{align}
	For a given form of $f({\cal R})$, we can solve the structure Eq.~\eqref{65} to write ${\cal R}$ as a function of $T$. Let us assume a polynomial form of 
	$f({\cal R}) = {\cal R}+\alpha {\cal R}^2$. This model has been  extensively studied in Palatini formalism, see for example  \cite{Barragan:2009sq, Szydlowski:2015fcq}. Then the solution of  structural Eq.~\eqref{65} has the following form
	\begin{align}
		{\cal R} = -\kappa^2 T.
	\end{align}
This model implies cosmological  power-law solutions   \cite{Goheer:2009ss} as
	\begin{align} \label{powerlaw}
		a(t) = a_0 \, t^m,
	\end{align}
	where $m>0$ is a real number. From Eqs.~\eqref{c1} and~\eqref{c2}, we have
	\begin{align}
		\rho_{\rm tot}(t) = \rho_{\rm m}(t)+\rho_{\rm r}(t)=\rho_{\rm m0}t^{-3m}+\rho_{\rm r0}t^{-4m}\,,
	\end{align}
	with $\rho_{\rm m0}$ and $\rho_{\rm r0}$ initial values.
	Therefore, we get
	\begin{align}
		2 \kappa^2 \, \tau^0_{\phantom{0}0} = \frac{6m(1-m)}{t^2} +\kappa^2 \rho_{\rm m0}t^{-3m} +6m (5-2m) \alpha \kappa^2 \rho_{\rm m0} t^{-3m-2} +\frac{108m^2 \alpha^2 \kappa^4 \rho_{\rm m0}t^{-6m-2}}{1+2m^2\kappa^2\rho_{\rm m0} t^{-3m}},
	\end{align}
	and
	\begin{align}
	\tau^0_{\phantom{0}0}+T^0_{\phantom{0}0} =-6\alpha m \rho_{\rm m0} t^{-(3m+2)}.
	\end{align}
	The total energy density of gravitational and non-gravitational fields is then
	\begin{equation}
	\sqrt{-g}(\tau^0_{\phantom{0}0}+T^0_{\phantom{0}0})=-6\alpha m \rho_{\rm m0} t^{-2},
	\end{equation}
	that goes to zero as the inverse square of  cosmic time.
	\section{The  metric formulation of $f(R)$ gravity} \label{metricsection}
	Similar considerations can be developed in metric formalism. The gravitational action of $f(R)$ gravity is given by
	\begin{align}
		{\cal S} = \frac{1}{2\kappa^2}\int {\rm d}^4x\, \sqrt{-g} \, f(R).
	\end{align}
	The variation with respect to the metric $g^{\mu\nu}$ and the coordinates $x^\mu$ gives
	\begin{align}
		\tilde{\delta} {\cal S} = &\frac{1}{2\kappa^2} \int {\rm d}^4x \, \sqrt{-g} \left[f_R(R) \, R_{\mu\nu}-\frac{1}{2}g_{\mu\nu}f(R)+g_{\mu\nu} \stackrel{\circ}{\square} f_R (R) - \stackrel{\circ}{\nabla}_\mu \stackrel{\circ}{\nabla}_\nu f_R (R) \right] \, \delta g^{\mu\nu} \nonumber \\ & +\int {\rm d}^4x \, \partial_\alpha \Bigg\{  \frac{\sqrt{-g}}{2\kappa^2} \Big[ \partial_\beta f_R(R) \, \left( g^{\eta \rho}g^{\alpha\beta}-g^{\alpha\eta}g^{\rho\beta}\right) \, \delta g_{\eta\rho} \nonumber \\& +f_R(R)\left[ (\stackrel{\circ}{\Gamma}{}^{\rho\eta\alpha}-\stackrel{\circ}{\Gamma}{}^{\eta\sigma}_{\phantom{\eta\sigma}{\sigma}} \, g^{\alpha \rho} ) \, \delta g_{\eta \rho} + (g^{\alpha\eta}g^{\tau\rho}-g^{\eta\rho}g^{\eta\tau}) \, \delta g_{\eta\rho,\tau} \right] \nonumber \\ & + f(R) \, \delta^\alpha_\lambda \, \delta x^\lambda \Big]\Bigg\},
	\end{align}
	where $f_R(R) := {\rm d}f(R) / {\rm d}R$ and the connection is Levi-Civita.
	Then imposing the action is stationary at a fixed $x$, and vanishing variation of both metric and its first derivatives at the boundary of the domain where the action is defined, we get the field equations in vacuum
	\begin{align}
		f_R(R)\, R_{\mu\nu}-\frac{1}{2}g_{\mu\nu}f(R) +\left(g_{\mu\nu} \stackrel{\circ}{\square}- \stackrel{\circ}{\nabla}_\mu \stackrel{\circ}{\nabla}_\nu\right) f_R(R) =0.
	\end{align}
	Then, for an infinitesimal transformation and considering that the local variation of the action vanishes, we get
	\begin{align}
		\partial_\sigma \left( \sqrt{-g} \, \tau^\sigma_{\phantom{\sigma}\lambda} \right) = 0,
	\end{align}
	where the gravitational energy-momentum pseudo-tensor in metric $f(R)$ is defined by
	\begin{align}
		2 \, \kappa^{2} \tau^\sigma_{\phantom{\sigma}\lambda} =& f(R) \, \delta^\sigma_\lambda-2 \, \partial_\beta f_R(R)\, g^{\eta[\rho} g^{\sigma]\beta} g_{\eta\rho , \lambda} - f_R(R)\, \left[\left(\stackrel{\circ}{\Gamma}{}^{\rho\eta\sigma}-\stackrel{\circ}{\Gamma}{}^{\eta\alpha}_{\phantom{\eta\alpha}\alpha} \, g^{\sigma\rho}\right)g_{\eta\rho , \lambda}+2g^{\sigma[\eta}g^{\tau]\rho}g_{\eta\rho,\tau\lambda}\right].
	\end{align}
	Considering the matter and using the definition of matter energy-momentum tensor, i.e. Eq.~\eqref{mEM}, we get the following field equations 
		\begin{align}
	f_R(R)\, R_{\mu\nu}-\frac{1}{2}g_{\mu\nu}f(R) +\left(g_{\mu\nu} \stackrel{\circ}{\square}- \stackrel{\circ}{\nabla}_\mu \stackrel{\circ}{\nabla}_\nu\right) f_R(R) =\kappa^2 T_{\mu\nu}.
	\end{align}
	Using the Bianchi Identities $\stackrel{\circ}{\nabla}{}^\mu G_{\mu\nu}=0$, the field equations result in the energy-momentum conservation, i.e. $\stackrel{\circ}{\nabla}{}^\mu T_{\mu\nu}=0$. Finally, for an infinitesimal rigid transformation,  we can obtain the local conservation of total energy-momentum density of matter and gravitation,
	\begin{align}
		\partial_\sigma\left[ \sqrt{-g} \left( \tau^\sigma_{\phantom{\sigma}\lambda} +T^\sigma_{\phantom{\sigma}\lambda}\right) \right]=0.
	\end{align}
	For flat FLRW spacetime, we can explicitly write the components $(00)$ of the gravitational energy-momentum  $\tau^\mu_{\phantom{\mu}\nu}$ and the
	the matter energy-momentum $T^\mu_{\phantom{\mu}\nu}$ as follows, respectively,
	\begin{align}
		\kappa^2\,	\tau^0_{\phantom{0}0} =& \frac{1}{2}f(R)-3\left(H^2+\dot{H}\right)\, f_R(R) +3H\dot{R} \, f_{RR}(R),
		\\ \label{f1}
		\kappa^2 T^0_{\phantom{0}0} =& - \frac{1}{2}f(R)+3\left(H^2+\dot{H}\right)\, f_R(R) - 3H\dot{R} \, f_{RR}(R).
	\end{align}
	Consequently,  the total energy of the gravitation and matter vanishes for FLRW spacetime, i.e.
	\begin{align}
	\tau^0_{\phantom{0}0}+T^0_{\phantom{0}0} =0.
	\end{align} 
	 We can assume a power-law evolution for  matter and radiation fluids  considering  Eq.~\eqref{powerlaw}.  We have
	 \begin{align} \tau^0_{\phantom{0}0} =&\rho_{\rm m}(t)+\rho_{\rm r}(t)\nonumber \\ =&\rho_{\rm m0} t^{-3m}+\rho_{\rm r0}t^{-4m}\,.
	 \end{align}
 The Ricci curvature scalar  is now
	\begin{equation}
		R = 12H^2+6\dot{H} = 6m(2m-1) t^{-2}\,.
	\end{equation}
	Then, the Friedmann equation  reduces into
	\begin{align}
		\frac{f_{RR}\, R^2}{(2m-1)}+\frac{m-1}{2(2m-1)} f_R \, R -\frac{1}{2} f +\kappa^2 \rho_{\rm m0} \left( \frac{R}{6m(2m-1)}\right)^{\frac{3}{2}m}+\kappa^2 \rho_{\rm r0} \left( \frac{R}{6m(2m-1)}\right)^{2m} =0.
	\end{align}
	From this equation, we get the specific form of $f(R)$ that results in a power law form.  It is
	\begin{align} \label{f}
		f(R) =& -\frac{4\kappa^2 \rho_{\rm m0} (2m-1)}{12m-11} \left( \frac{R}{6m(2m-1)}\right)^{\frac{3}{2}m} -\frac{2\kappa^2 \rho_{\rm r0} (2m-1)}{10m^2-8m+1} \left( \frac{R}{6m(2m-1)}\right)^{2m} 
		\nonumber \\
		&+C_1 R^{\frac{3}{4}-\frac{m}{4}-\frac{1}{4}\sqrt{m^2+10m+1}} +C_2 R^{\frac{3}{4}-\frac{m}{4}+\frac{1}{4}\sqrt{m^2+10m+1}}.
	\end{align}
	For $m=2/3$ and $\rho_{\rm r0}/\rho_{\rm m0} \ll 1$, we reduce to $f(R)\sim R$ and GR is recovered.
	
	\section{Conclusions}\label{conclusions}
	We have 
	calculated the gravitational energy density  in the context of $f(R)$ gravity for both  Palatini and metric approaches. This quantity, being described by an affine tensor, a pseudo-tensor,  is not localizable.  The main result is that it is possible to obtain two general expressions for the gravitational energy-momentum $\tau^\mu_{\phantom{\mu}\nu}$  in both Palatini and metric approaches.  Considering a flat FLRW spacetime, we found that the energy density complex vanishes in the metric formulation of $f(R)$ gravity. Similar results have been achieved in  GR  and in teleparallel gravity~\cite{Abedi:2018eyz, Faraoni:2002vw, Johri:1995gh, Vargas:2003ak, Rosen:1994vj, Abedi:2015cya}. On the other hand,  the complex is  non-vanishing in the $f({\cal R})$ Palatini formulation except for  the case   $f({\cal R}) = {\cal R}$.   This fact could be relevant in discriminating among theories, in particular in view of gravitational radiation where the energy-momentum  complex can play an important role.  As reported in \cite{Capozziello:2017xla},  further modes of gravitational radiation, beyond the standard two modes of GR, are related to the  gravitational stress-energy pseudo-tensor emerging from alternative and extended theories of gravity. 	In this perspective, a detailed analysis of this pseudo-tensor is fundamental for gravitational waves and multi-messenger astronomy.
	
\section*{Acknowledgments}
SC and MC acknowledge the Istituto Nazionale di Fisica Nucleare (INFN) Sez. di Napoli (Iniziative Specifiche QGSKY and MOONLIGHT2) for the support.

\end{document}